\documentclass[traditabstract]{aa} 
\usepackage{graphicx}
\usepackage{subfig}
\usepackage{txfonts}
\usepackage{natbib}
\newcommand{\Hi}{\textup{H\,{\mdseries\textsc{i}}}}
\newcommand{\HI}{\textup{H\,{\mdseries\textsc{i}}}~}

\newcommand{\Msun}{{M$_{\odot}$}}

\def\kms{{km~s$^{-1}$}}

\def\degr{\hbox{$^\circ$}}

\def\fdegr{\hbox{$.\!\!^\circ$}}

\begin{document}
   \title{An \HI absorbing circumnuclear disk in Cygnus A}

   \subtitle{}

   \author{C. Struve\inst{1} 
   \and 
   J.E. Conway\inst{2}
   }
   
  \institute{Netherlands Foundation for Research in Astronomy,
             Postbus 2, 7990~AA, Dwingeloo, The Netherlands \\
            Kapteyn Institute, University of Groningen,
             Landleven 12, 9747~AD, Groningen, The Netherlands \\
            \email {struve@astron.nl},   
          \and
          Onsala Space Observatory, SE-439 92 Onsala, Sweden \\
\email{john.conway@chalmers.se}
}


   \date{Received 30 October 2009; accepted 19 December 2009}

\abstract {We present Very Long Baseline Array (VLBA) \HI absorption observations of the core region of the powerful radio galaxy Cygnus~A. These data show both broad (FWHM~$= 231\pm 21$\kms ) and narrow (FWHM~$ <30 $\kms ) velocity width absorption components. The broad velocity absorption shows high opacity on the counter-jet, low opacity against the core and no absorption on the jet side. We argue that these results are  most naturally explained by a circumnuclear \HI absorbing disk orientated roughly perpendicular to the jet axis. We estimate that the \HI absorbing gas lies at a radius of $\sim 80$~pc has a scale height of about 20~pc,  density $n>10^4$~cm$^{-3}$ and total column density in the range $10^{23}-10^{24}$~cm$^{-2}$. Models in which the \HI absorption is primarily from an  atomic or a molecular gas phase can both fit our data.  Modelling taking into account the effective beam shows that the broad  \HI absorbing gas component does not cover the radio core in Cygnus~A and therefore does not contribute to the gas column that blocks our view of the hidden quasar nucleus. If however Cygnus~A were observed from a different direction, disk gas on $\sim100$pc radius scales would contribute significantly to the nuclear column density, implying that   in some radio galaxies gas on these scales may contribute to the obscuration of the central engine. We argue that the circumnuclear torus in Cygnus~A contains too little mass to power the AGN over $>10^7$~yr  but that material in the outer \HI absorbing gas disk can  provide a reservoir to fuel the AGN and replenish torus clouds. The second narrow \HI absorption component is significantly redshifted (by $186$\kms ) with respect to the systemic velocity and probably traces infalling gas which will ultimately fuel the source. This component could arise either within a tidal tail structure associated with a recent (minor) merger or be associated with an observed infalling giant molecular cloud.}


   \authorrunning{Struve \& Conway}

   \maketitle
%

\section{Introduction}

Circumnuclear obscuring tori/disks  are an essential component of unified schemes of active galactic nuclei \citep[e.g.][]{antonucci93, tadhunter08}. 
Recently there has been much progress in obtaining direct evidence for such structures especially in Seyfert  luminosity objects.  
This evidence includes modelling of  the IR SEDs from AGN heated  dust in clumpy tori \citep{nenkova08} and direct IR interferometric imaging  of this dust on 1~pc - 10~pc scales \citep{jaffe04, tristram09}.

 On larger scales adaptive optics IR observations of molecular hydrogen lines \citep{hicks09} have revealed geometrically thick gas at radii 30~pc in Seyferts. Additionally millimetre interferometry also detects 
 molecular gas in emission on scales $r = 70$~pc \citep[e.g.][]{schinnerer00} albeit in more flattened disk-like structures. Such outer disk structures may be 
 continuous with inner obscuring tori and provide both the fuel and a conduit for feeding the central engine. The relationship between these circumnuclear disks and obscuring   tori is  however far from clear. 
 
Radio observations  provide another means to study  the circumnuclear gas environment. This can for instance be achieved by VLBI observations of maser emission from molecular gas \citep{lo05}, free-free absorption from ionised gas and absorption from atomic gas (\Hi ). Examples of the use of the latter two tracers include observations of NGC 1275 \citep[e.g.][]{vermeulen94},  Centaurus~A \citep[e.g.][]{jones96,morganti08}, Hydra~A \citep[e.g.][]{taylor96}, NGC~4261 \citep{vanlangevelde00,jones01} and $1946+708$ \citep{peck01}.  Because of their high spatial resolution such radio observations are especially suitable
for studying circumnuclear  obscuring matter in powerful narrow line radio galaxies which are expected \citep[see][]{tadhunter08} to be unified via orientation with radio-loud quasars. There is strong evidence from
X-ray observations \citep{hardcastle09} for the expected obscuration by large column densities in the former objects, however they are usually too distant and faint for optical and IR observations to directly 
observe  the circumnuclear gas.

A prime target for studies of circumnuclear gas  in a  luminous 'hidden quasar'  radio galaxy is the closest  Fanaroff-Riley \citep{fanaroff74} type II (FR-II) radio-galaxy Cygnus~A.
Spectropolarmetric observations of this source  which revealed a hidden BLR in scattered light \citep{ogle97} were a major milestone in the general acceptance  of the orientation unification scheme for powerful radio galaxies and radio-loud quasars. Further evidence for shadowing from a central torus comes from the bi-cones observed in both optical emission lines  \citep{jackson98} and IR continuum  \citep{tadhunter99}. The sharpness of the edges of these bi-cones  suggest that the inner face of any torus occurs at  radii $<50$~pc from the central engine  \citep{tadhunter08}.  \citet{tadhunter03} have from optical/IR emission line  
observations measured  a rotation curve from gas rotating  around the bi-cone/radio-jet axis at $r \approx 300 -1000$~pc allowing a central black hole mass to be estimated. These observations may trace the outer parts of a circumnuclear gas structure which connects with the inner obscuring torus. The ultimate origin of this  material could be related to the merger activity detected in Cygnus~A \citep{canalizo03}. 

The central radio core and inner jets of Cygnus A are relatively bright from millimetre to centimetre wavelengths allowing searches in absorption to constrain circumnuclear gas properties on small ($<100$~pc) scales. Molecular absorption observations so far give ambiguous results with only upper limits or marginal detections being reported  \citep[see e.g.][]{barvainis94,fuente00,salome03,impellizzeri06}. VLBI observations by \citet{krichbaum98} have however found  evidence for ionised circumnuclear gas  on scales $<20$~pc via the detection of free-free absorption toward the counter-jet. Additionally \citet{conway95} detected broad \HI absorption toward the core in VLA observation  which were interpreted in terms of a circumnuclear disk/torus model, with the \HI absorption either 
tracing the small atomic fraction of a mainly molecular medium or a purely atomic structure. To better constrain the scale and geometry of this \HI absorbing gas we have performed high resolution NRAO Very Long Baseline Array (VLBA\footnote{The VLBA is operated by the National Radio Astronomy Observatory which is a facility of the National Science Foundation operated under co-operative agreement by Associated Universities, Inc.}) \HI absorption observations. A short report on a initial reduction of this data was given by  \citet{conway99}, this present paper presents a fuller re-analysis of the data. The organisation of this paper is as follows, in Sect.~2 we describe the observations while 
Sect.~3 presents the observational results including modelling of  opacity profiles along the source. A discussion of the results is given in Sect.~4 and a summary in Sect.~5. 
At the redshift of Cygnus~A ($z=0.056$) for cosmologies with H$_{\rm{o}} = 73$ \kms ~1~mas corresponds to approximately 1~pc,  a scaling which we adopt throughout this paper. All total recession velocities quoted are heliocentric (optical definition).

%
\begin{figure*}
\centering
\includegraphics[width=0.90\textwidth]{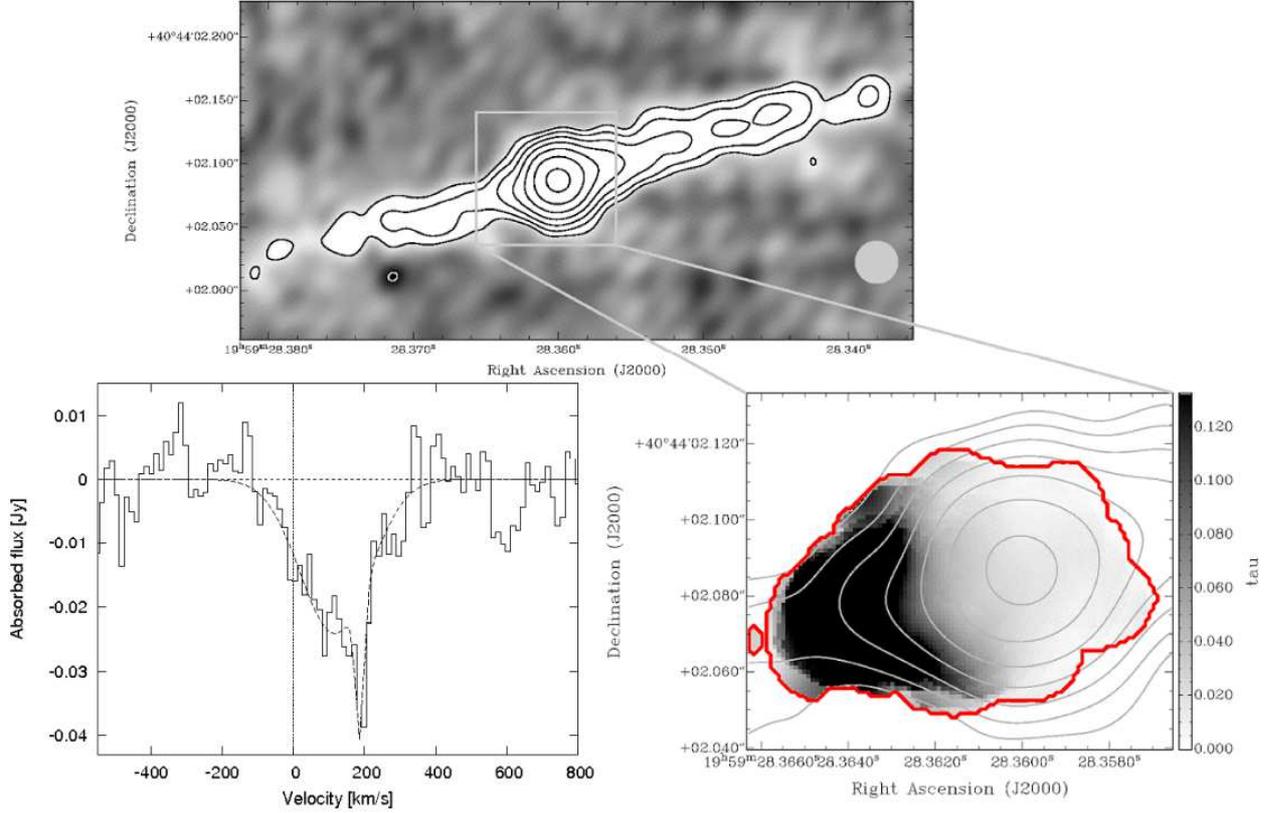}
\caption{Top panel: Continuum image at 1340~MHz. The lowest contour is at 2~mJy~beam$^{-1}$ with subsequent contours increasing by  factors of 2. The effective beam FWHM (see Sect.~\ref{se:dataobs}) is indicated in the lower right corner. Bottom left panel: Integrated absorption spectrum from the blanked cube with the velocities shifted to the rest frame of Cygnus~A (see Sect.~4.1). The dashed line shows the two component Gaussian fit. Bottom right panel: Contours show continuum. Grayscale shows the mean opacity over the rest frame velocity range $-80$ to $+170$~\kms . The thick dark line shows the un-blanked region over which the integrated absorption spectra (shown in the bottom left panel) is calculated.  
 }
\label{cont}
\end{figure*}
%
%
\section{Observations} \label{se:dataobs}

Observations were performed on August 31st 1995 using the ten stations of the VLBA plus the phased VLA. Two IFs (left and right circular polarisation) with a bandwidth of 12.5~MHz and 256 channels were centred at the frequency of the previously observed \HI absorption \citep[$\sim1340$~MHz,][]{conway95}. A number of bright compact sources  were observed as fringe finders and for bandpass calibration. The data were correlated in Socorro, USA. A standard data reduction using AIPS was performed including fringe fitting, calibration and flagging of the data. Cygnus~A lies at a relatively low galactic latitude ($b= 5$\fdegr 7) and the effects of interstellar scattering are significant at VLBI resolution. For this reason four antennas which participated only in long baselines, namely  Brewster, Hancock, Mauna Kea and St. Croix  showed no fringes to Cygnus A  and so were deleted from the subsequent  data analysis.  Initial amplitude calibration was accomplished for VLBA antennas using  the recorded system temperature values (which took into account the dominant contribution to the noise from the lobes of Cygnus~A). For the phased VLA the calibrator  T$_{\rm{ant}}$/T$_{\rm{sys}}$ values were used with a correction applied to take into account the system noise contribution from Cygnus A.  Offsets in the VLA amplitude calibration scale were then corrected by comparing the correlated Cygnus A  flux densities on long baselines to respectively the VLA and Pie Town. Initial continuum images were made via iterative  phase self-calibration/deconvolution  starting from a point source. A couple of cycles of amplitude  and phase self-calibration were performed at the very end to obtain a noise limited continuum map. 

Extensive experiments were carried out to determine the optimum uv weighting which gave the best combination of sensitivity and resolution for both continuum and spectral line. Final images were made using uniform weighting with robustness factor 0.5; giving rise to an almost circular dirty beam with main lobe width FWHM $\approx 25$~mas. Since the uv point  weights also took into account the sensitivity of each baseline (by weighting by 1/noise variance)  the final continuum 
and spectral line images are dominated by the baselines to the VLA. CLEAN images  were restored with a circular FWHM 25~mas Gaussian; however the effective image resolution is less than this because of the effects of foreground interstellar scattering. The observed effective resolution (i.e. the clean beam convolved with the interstellar scattering) is 32.7~mas, as determined by measuring the FWHM of the continuum profile perpendicular to the jet at the core position - implying an interstellar scattering contribution of 21~mas.

The resulting continuum image (see Fig.~\ref{cont}) has rms noise of 0.43~mJy~beam$^{-1}$ and shows besides the unresolved core a jet and counter-jet structure (see Sect.~3). In  principal it is possible that  the counter-jet structure could be an artefact of phase self-calibration starting from a point source. To check this possibility  we re-mapped the data only allowing flux density on the jet side to be included in the model in the initial cycles, but in all cases emission on the counter-jet side remained. The existence of a counter-jet is also confirmed by other observations \citep[e.g.][]{krichbaum98}. In making our final continuum and spectral line images we used  data self-calibrated against continuum models including both jet and counter-jet 
emission. Before making our spectral line cube we removed the continuum contribution using the AIPS task UVLIN and then imaged using the same weighting as used for our continuum image and the same CLEAN restoring beam.
In order to increase sensitivity when creating our spectral line cube  the uv data were averaged in frequency to give a final channel separation  equivalent to 14.0~\kms ~in velocity  or a velocity resolution of 28.0~\kms ~after Hanning smoothing. The noise achieved in the final line cube was $\sigma_{\rm{rms}}=2.36$~mJy~beam$^{-1}$ per channel.\\

\section{Results}

\subsection{Continuum image}
\label{contres} 

The 1340~MHz continuum image (Fig.~\ref{cont}, top panel) is very similar to the 1660~MHz image shown by \citet{krichbaum98}. We clearly detect the unresolved core, the jet and the weaker counter-jet (SE of the core). The PA of the jet is 105\degr $\pm 2$\degr , in agreement with VLBI observations at higher frequencies \citep{krichbaum98} and the kpc-size jet structure \citep{carilli91}. The continuum peak has brightness 0.16~Jy~beam$^{-1}$ and  the total continuum flux recovered in our observations is 0.50~Jy.

\subsection{\HI absorption}
\label{HIabsres}

An integrated absorption spectrum over our spectral line cube is shown in Fig~\ref{cont}, bottom left. Broad ($\Delta v=456$~\kms ) \HI absorption is detected in the velocity range from 16679 to 17135~\kms  with the peak being located at $v=17002$~\kms ~($z=0.05667$). 
The integrated spectrum is well fitted by two Gaussian components, yielding centroid velocities of $16916\pm 10$~\kms ~($z=0.05639\pm 0.00003$) and $16986\pm 5$~\kms ~($z=0.05662\pm 0.00002$), with FWHM$= 231\pm 21$ and $29\pm 10$~\kms ~respectively (after correcting the line widths to the rest frame of the host galaxy). The second, narrow component has a FWHM velocity width similar to our velocity resolution and hence we consider this velocity width as an upper limit of the true line width. The presence of two components in the integrated spectrum suggests that the \HI absorption consists of two different overlapping structures. Despite small differences in the flux scale (for continuum and absorption spectrum)  the VLBA data agree in absorption width, profile shape and radial velocity with the VLA A- and B-array observations of \citet{conway95}. These authors found slightly different Gaussian fits for the absorption spectrum, their data however potentially suffers from bandpass calibration and continuum subtraction problems (since two IFs --- only slightly overlapping in velocity --- had to be used to achieve sufficient spectral resolution and velocity coverage). Given these issues we conclude that both data sets agree within the noise of both observations so that we have recovered the full absorption seen by the VLA. The difference seen in flux scale between our VLBI and the published VLA observations are likely due to inaccuracies in amplitude calibrations of the VLA. This instrument, unlike VLBI, did not record the antenna system temperatures  which are greatly enhanced due to the presence of the bright radio lobes of Cygnus~A  in the primary beam of each antenna, complicating the amplitude calibration.
 
\begin{figure}
\centering
\includegraphics[width=0.47\textwidth]{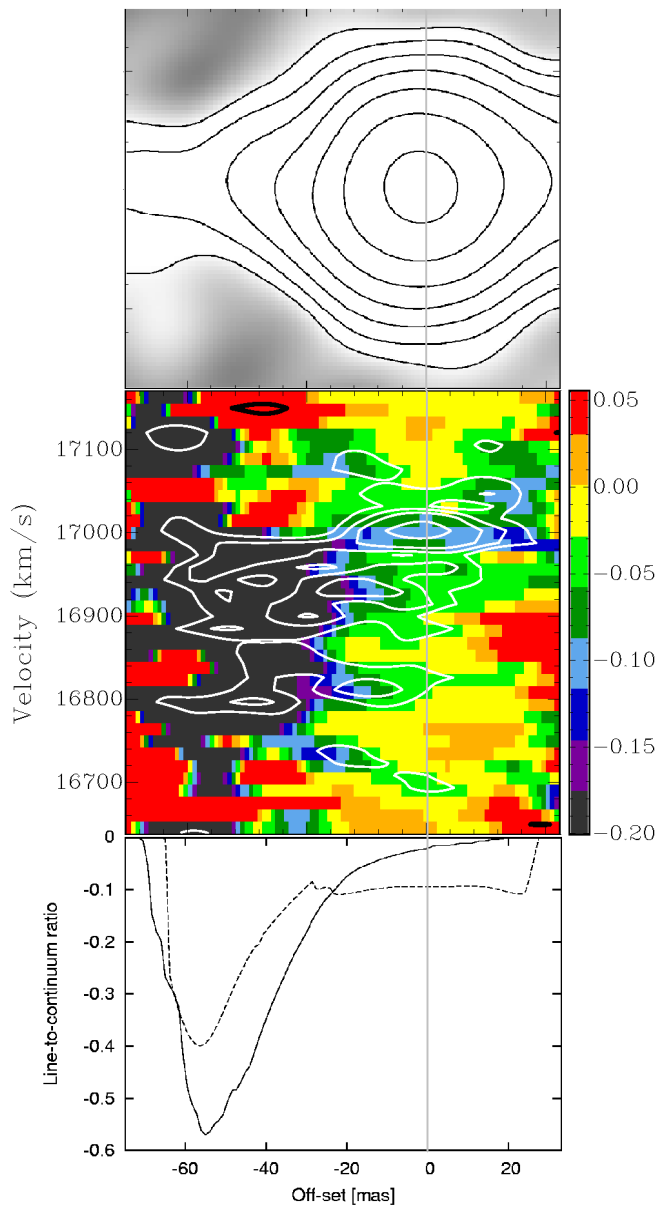}
\caption{Top panel: Radio continuum image of the core region rotated such that the jet is pointing to the right and the counter-jet to the left. Contour levels are the same as in Fig.~\ref{cont}. Middle panel: Position-velocity diagram along the radio axis. Grey scale shows the line-to-continuum ratio. Over-plotted in contours is the absorbed flux density in ~mJy~beam$^{-1}$. Contour levels are -15, -10, -7.5, -5.0 (white) and 5.0~mJy~beam$^{-1}$ (black). Bottom panel: Velocity averaged line-to-continuum ratio along the source for two different velocity ranges. The solid line is for a velocity range dominated  by the broad velocity component ($16720-16970$~\kms ), the dashed line for a velocity range ($16971-17012$~\kms ) centred on the narrow absorption system velocity.}
\label{opacity.pv}
\end{figure}

Inspection of the data cube shows that the absorption is spatially resolved and is  detectable over 95~mas in angle along the radio axis (i.e. $\sim3$ effective beam FWHMs). Below the second contour of the continuum image 
it is not possible to constrain the \HI absorption because the background is too weak.
The deepest absorption measured in mJy is toward the counter-jet and unresolved core but the highest opacities occur for the broad velocity component on the counter-jet side (see Fig.~\ref{cont}, bottom right). We find no indication of changes in spectral line absorption profiles in directions perpendicular to  the jet axis - which is as expected given the small jet width \citep{krichbaum98} compared to our effective resolution; this means we need to only consider the spectral profile as a function of position along the jet axis as shown in Fig.~\ref{opacity.pv}. In this figure the top panel shows the rotated continuum image while the middle panel shows the absorbed flux density (contours) and absorbed flux/continuum ratio (colours; note: saturated over 0.2) versus velocity and position along the jet.

The highest contours in the middle panel of Fig.~\ref{opacity.pv} belong to the narrow velocity absorption component at recession velocity 16986~\kms ~seen against the core. At the same velocity a  slight extension on the jet-side is detectable, consistent with having the same  $\tau \approx 0.1$ opacity as seen on the core. Because of the rapid fall-off of continuum intensity further along the jet and counter-jet further information about the spatial  distribution of the narrow velocity component is limited. That is to say that the underlying continuum is not strong enough to detect the narrow velocity absorption component even if the opacity remains 
$\tau \approx 0.1$.

The broad absorption component in Fig.~\ref{opacity.pv}, middle panel (seen at velocities around 16900 \kms)  is detected from the core position out to 65~mas along the counter-jet. Over this spatial range the broad component has absorbed flux densities (contour levels) which stay almost constant; given the rapid fall off in background continuum this corresponds to a rapid increase in line-to-continuum ratio (colours). Over the range of position from -50~mas to $-65$~mas spectra taken show the absorption to be flat bottomed and the absorbed flux density comparable with the continuum, both implying \HI opacities $\tau \gg 1$. The absorption shows an apparent sharp decrease beyond $-65$~mas, though in a region with very weak continuum (we discuss further the reality of this decrease in Sect 3.3). The broad absorption also apparently has a  small but significant opacity ($\tau \approx 0.03$) at the position of the core, however, this may be due only to the limited spatial resolution causing "leakage" of absorption onto the core position. Quantitative estimates of broad line opacity along the source are made in Sect.~\ref{se:modelbroad} where the above two points are addressed.

Reinforcing the above description of the two velocity components the bottom panel of Fig.~\ref{opacity.pv} shows the line to continuum ratio averaged over different velocity ranges versus position. The solid line shows average line-to-continuum  for velocity ranges where the broad component dominates, while the dashed line is for a velocity range where the narrow absorption is most dominant. The broad component shows rapidly increasing absorption along the counter-jet reaching a peak velocity averaged line-to-continuum ratio of almost 0.6 at $-55$~mas (with peak ratios within the velocity profile at this position in fact reaching up to 1 and beyond). The second profile (dashed line) for the velocity range where the narrow velocity component normally dominates  has approximately the same mean opacity on the core and jet. On the counter-jet side the average opacity over this velocity range increases, this is however consistent with the 
narrow absorption component having the  same opacity as  on the core and jet-side but with the average line-to-continuum ratio becoming dominated by contamination from the high velocity wings of the broad velocity component.

%
%
%

\subsection{Modelling spatial variations in broad line opacity }
\label{se:modelbroad}

The increase in line-to-continuum ratio of the broad absorption seen on the counter-jet side (see Sect.~3.2 and Fig.~\ref{opacity.pv}) implies a rapid increase in opacity. As noted in Sect.~3.2, at the position of maximum absorption the spectral profiles are flat-bottomed and saturated implying large opacity. To more accurately convert the line-to-continuum ratios versus velocity to peak opacity, and therefore estimate \HI column density variations, we have carried out a detailed modelling of the data. In this modelling we describe both continuum and peak opacity estimates with a set of seven point sources separated by 15~mas along the jet/counter-jet axis (i.e. just under half an effective beam FWHM width). Additionally by taking into account the effective beam width this modelling provides a modest super-resolution of the data, important because interstellar scattering limits our spatial resolution (Sect.~2). Specifically, we are interested in the question of whether the weak broad absorption apparently seen at the core position (see Fig.~\ref{cont} bottom right panel) is real or whether it can be explained by the combination of very strong absorption on the counter-jet side combined with limited spatial resolution.

In our modelling we first made estimates for the underlying continuum profile for each model point source (referred to as a  "pixel") where there was detectable absorption.  
Continuum intensities were varied at each of these seven pixels until after convolution with the effective restoring beam the model continuum profile versus position fitted the observations.  In a similar way we estimated for each pixel the absorbed line area (mJy~\kms ) at velocities in the range 16720 - 16970~\kms ~(where the broad absorption component dominates). Pixel values were again adjusted such that after convolution by the effective beam the model fitted the observations. Finally an estimate was made at each pixel of the peak \HI opacity versus velocity by combining the pixel-based continuum and spectral line absorption estimates. In estimating this quantity we assumed the  broad-component opacity spectrum was Gaussian with fixed velocity centroid and fixed FWHM.

\begin{figure}
\centering
\includegraphics[width=0.33\textwidth, angle=270]{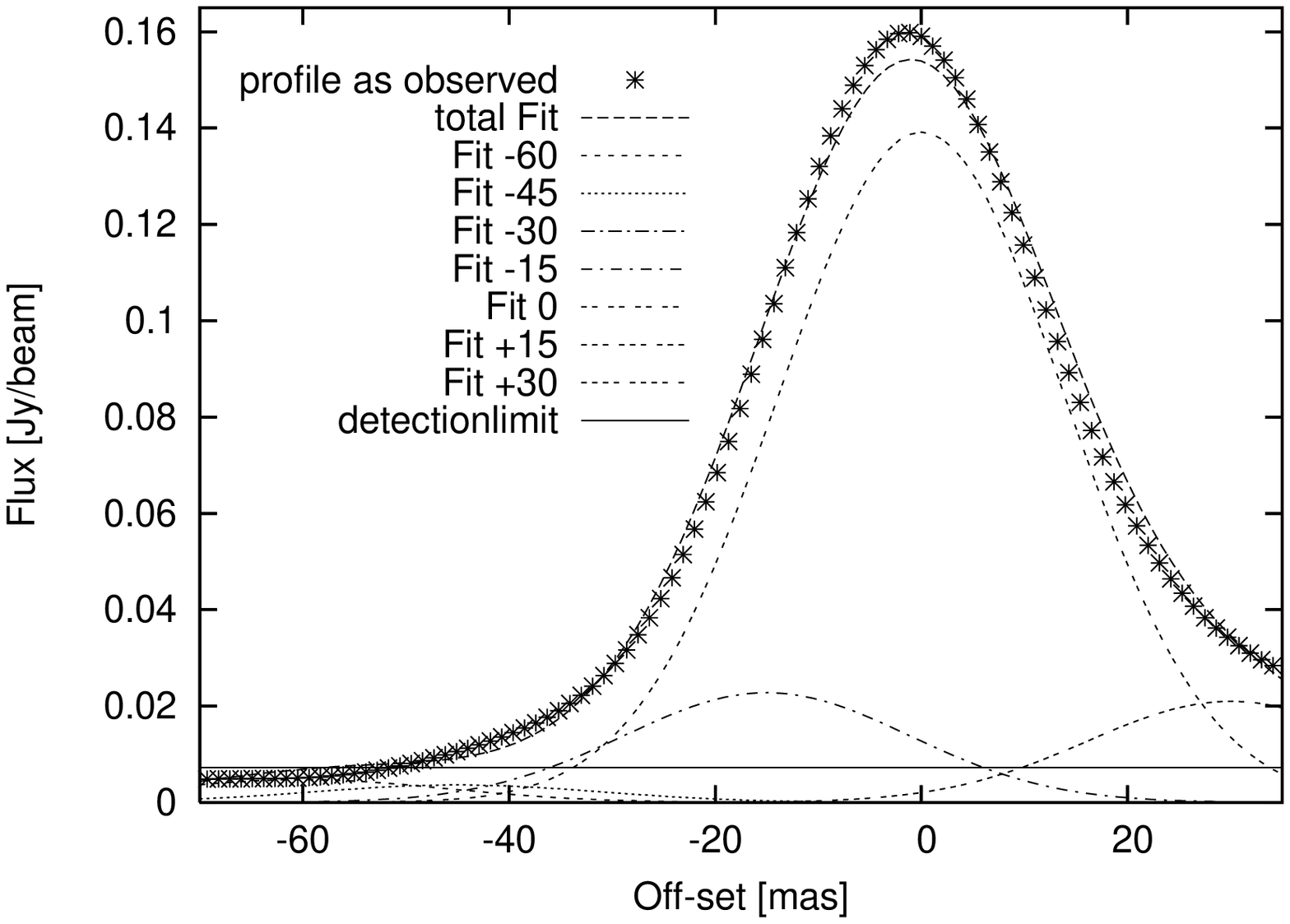}
\includegraphics[width=0.33\textwidth, angle=270]{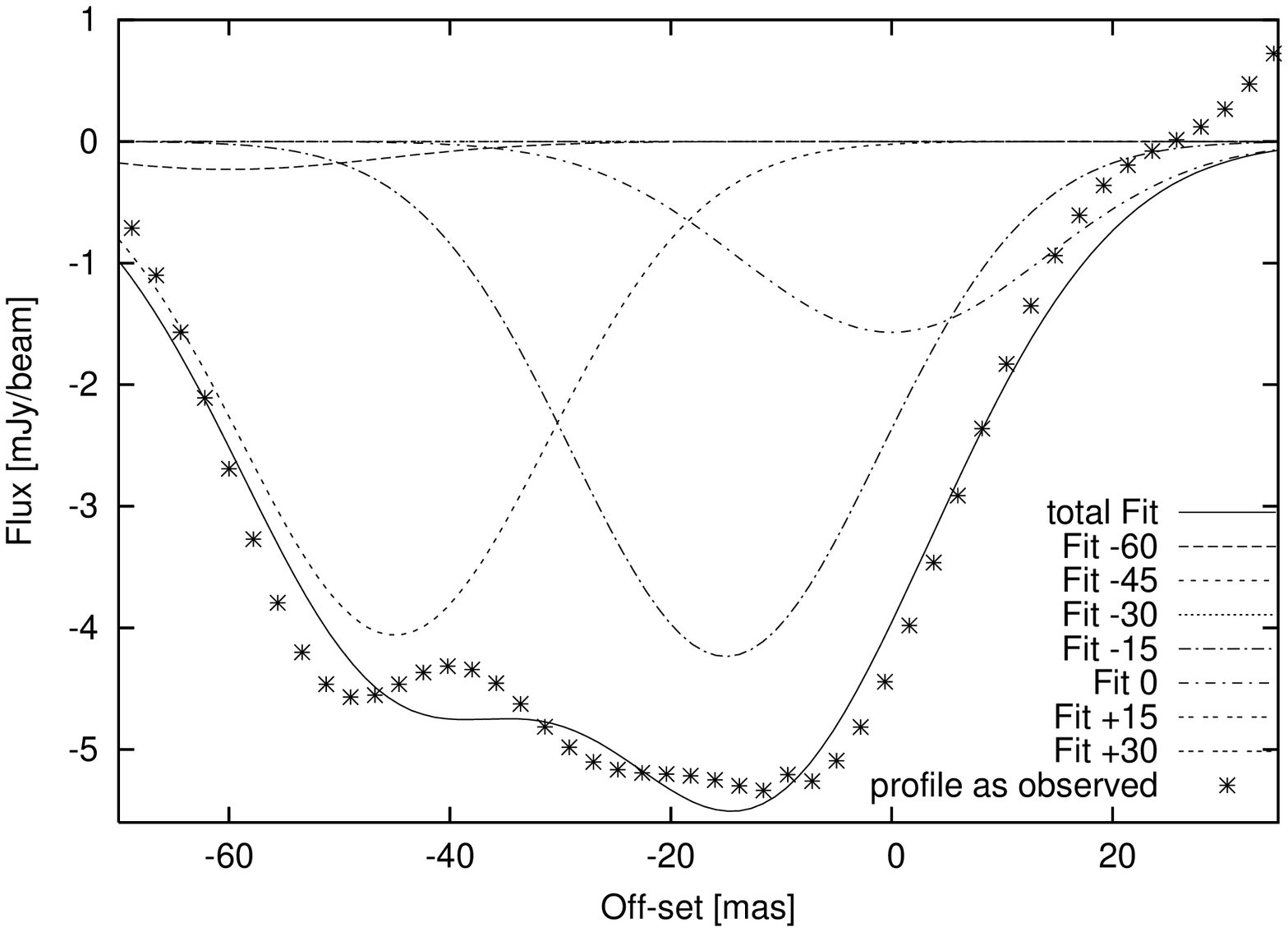}
\caption{Modelling of continuum and broad absorption along the jet axis (see Sect. 3.3). Top panel: Comparison of model fit and data  for the observed continuum profile found by adjusting pixel values while assuming an effective Gaussian beam of FWHM of 32.7~mas. Bottom panel: Comparison of model and data for average absorbed line flux over the broad component velocity range from $16720 - 16970$~\kms . In both plots negative offsets correspond to counter-jet, positive offsets to jet side.}
\label{cont.fit}
\end{figure}

\begin{figure}
\centering
\includegraphics[width=0.33\textwidth, angle=270]{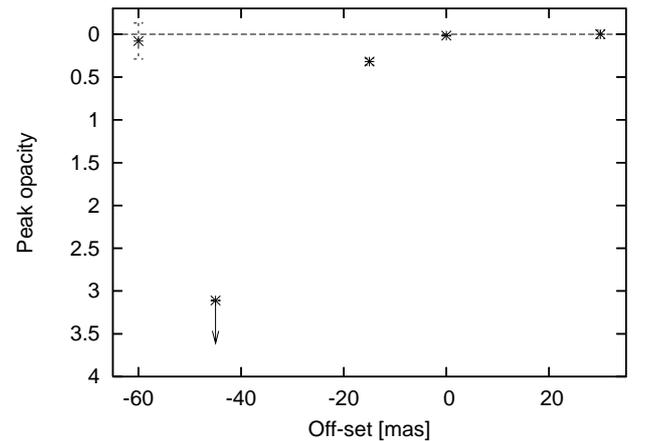}
\caption{Estimates of broad absorption opacity at line centre versus position 
along the radio axis after taking  into account beam effects (see Sect. 3.3). The opacities and errors plotted
are taken from Tab.~1.}
\label{decon.ratio}
\end{figure}

The resulting fits reproduce (see Fig.~\ref{cont.fit}), to first order, both the observed continuum and the absorbed flux profiles along the source. Deviations of the fits from the data are likely due to the background continuum being more complicated \citep[see][]{krichbaum98} than our parameterisation of it. Table.~1 gives our results and the line centre peak opacity as plotted in Fig.~\ref{decon.ratio}. Note that no opacity estimate  is plotted at positions -30 and $+15$~mas because the fitted continuum intensity at these positions was zero. We estimate the error of the line-to-continuum ratio (Tab.~1, column 4) assuming it is dominated by the rms noise in the line data.  At  position $r=-45$~mas the lower limit  on line centre opacity is set by subtracting  2$\sigma$ from the line-to-continuum ratio.

The peak opacity (column 5) takes into account that part of the absorption spectrum is flat bottomed and the attached error is calculated based on a $1\sigma_{\rm{rms}}$ uncertainty in our line data. Our results are consistent with no broad \HI absorption on the jet-side and any opacity against the core being very low ($0.016$). Moving outward from the core along the counter-jet, the opacity increases and peaks at 45~mas ($N_{\Hi}=1.6\cdot 10^{21}$T$_{\rm{spin}}$) before it sharply decreases to 0.078 at 60~mas from the core. Based on uncertainty estimates of the absorbed flux this drop in opacity is real and not an observational artefact due to insufficient signal-to-noise.

%
%
%
%
\begin{table}
 \caption{Results of  modelling  of continuum and broad  \HI absorption 
 profiles along the jet axis (see Sect- 3.3).}
\centering
\begin{tabular}{r r c c c}

\hline\hline
r & F$_{C}$  & F$_{L}$  & F$_{L}$F$_{C}^{-1}$  & $\tau_0$  \\
(mas)  & (mJy) & (mJy)   &    &  \\
\hline
-60 & 4.73 & 0.23 & $0.05 \pm 0.12$ & $0.080 \pm 0.210$\\
-45 & 3.62 & 4.06 & $1.12 \pm 0.15$ & $>3.110$\\
-30 & 0.02 & 0.00 & 0.00 & -\\
-15 & 22.78 & 4.24 & $0.19 \pm 0.02$ & $0.320 \pm 0.040$\\
0 & 139.16 & 1.57 & $0.01 \pm 0.004$ & $0.016 \pm 0.006$\\
15 & 0.00 & 0.00 & - & -\\
30 & 20.98 & 0.00 & $0.00 \pm 0.03$ & $0 \pm 0.0478$\\
\hline
\label{table}
\end{tabular}
\caption*{Column 1 give the distance from the core of the fitted pixel, negative offsets correspond to positions along the counter-jet, positive to positions on the jet side. Columns 2 to 4 give respectively the  fitted continuum flux density, \HI absorbed flux density averaged over fitted  frequency range and the line to continuum ratio. Column 5 gives the modelled opacity at line centre.}
\end{table}
%
%
%
%

\section{Discussion}

\subsection{A disk geometry for the broad absorption component}

The fact that the broad velocity width absorption is seen at high opacity only against the counter-jet and not against the jet is similar to the situation found for other powerful cores of radio galaxies observed in \HI absorption, for instance in the FR-I source NGC~4261 \citep{vanlangevelde00}. The most natural explanation for these observations is that the \HI absorption lies in a flattened structure roughly in a plane perpendicular to the radio axis (i.e. a circumnuclear disk). Alternative explanations involving foreground clouds in the ISM of the host galaxy seem unlikely, requiring within elliptical hosts a large covering factor of 10pc-100pc sized clouds with uncharacteristically large internal velocity dispersion (of order 300~\kms ).

Converting our observations of the location and distribution of  \HI absorption along the counter-jet to an exact of radius in the disk depends on the precise orientation of the disk. Over the range of plausible disk orientations however this radius varies over a fairly narrow range as can be seen by examining two limits, first when the disk is close to edge-on and when at its maximum plausible tilt.  In the former case the \HI profile width  in Fig.~\ref{decon.ratio}~ measures the scale height of the disk. Since we know the central black hole mass  \citep[M$_{\rm{BH}}=2.5\cdot 10^9$~\Msun ,][]{tadhunter03} the radius at which the \HI absorption occurs can be calculated \citep[following][]{krolik88} via:

\begin{equation}
\frac{\Delta h}{r}\sim\frac{\Delta v}{v_{\rm{rot}}(r)}
\end{equation}

\noindent where $\Delta h$ is the disk scale height, $r$ the radius from the core, $\Delta v=98.5$~\kms ~the velocity dispersion (estimated from the Gaussian fit in Sect.~3.3) and $v_{\rm{rot}}(r)$  the Keplerian rotation velocity at radius $r$ from the  central black hole. Assuming $\Delta h = 25$~pc and solving for $r$ we obtain $r=88$~pc. In the other limit we consider a  maximally inclined disk. Based on VLBI observations of the jet to counter-jet brightness ratio of the parsec scale jet \citep{krichbaum98} the jet axis is orientated at an angle $\theta >80$\degr ~to the line of sight.  Based on measured misalignments of larger scales  \citep{tadhunter03} we estimate a  maximum misalignment between disk and radio axes of $30$\degr - which gives a maximally inclined disk which is $40$\degr ~from edge-on. The peak \HI opacity is observed to occur at projected distance 45~pc along the counter-jet which  when de-projected corresponds to a radius of 70~pc. Using again eq.~1 this gives a scale height of $\sim 18$~pc. Based on the above two limits we will assume in the following that the \HI peaks at a radius 80~pc from the black hole and has a scale height of about 20~pc. The resulting opening angle of the circumnuclear \HI disk is $\sim14$\degr ~which is similar to what is found in other sources \citep[e.g. in NGC~4261][]{vanlangevelde00}.

An important question in considering the feasibility of the disk hypothesis is the velocity of the broad absorption relative to the systemic velocity of the galactic nucleus. For a disk which is perfectly normal to the jet axis these velocities will be the same.  Our best estimate of the systemic velocity 
we take to be $z=0.05600\pm 0.00008$ ($=16800\pm24$~\kms ) which is the mean of six published optical/IR emission line estimates  \citep[see Table.~1 in][]{tadhunter03}.
The centroid of the broad component at 16916~\kms ~is therefore 116~\kms~  beyond the mean systemic velocity. The observed offset however can be accommodated if there were a fairly modest misalignment between the disk axis and the jet. VLBI observations constrain the jet axis to be within 10\degr ~of the sky plane. For such an orientation misalignments between projected jet axis and disk axis will be similar to intrinsic misalignments. Given the arguments above that the \HI absorption occurs at  $\sim 80$~pc radius and given the estimated central black hole mass of \citep[M$_{\rm{BH}}=2.5\cdot 10^9$~\Msun ,][]{tadhunter03}  the orbital velocity is $v_{\rm{rot}}(r=80\rm{pc})= 367$~\kms. Given this orbital velocity a misalignment of only 21\degr ~is sufficient to explain the difference between the \HI centroid and systemic velocities.

\subsection{Physical properties of the broad absorption component gas}

According to \citet{maloney96a} the physical state of  gas around an AGN  is controlled by an ionisation parameter determined from the ratio of the hard X-ray photon flux to local gas density.  At  a given radius $r$ this ionisation parameter equals
\begin{equation}
\xi_{\rm{eff}}=L_{\rm{X}}\cdot n^{-1}\cdot r^{-2}\cdot N_{22}^{-0.9}
\end{equation}
where $L_{\rm{X}}$ is  the luminosity in  $>2$~keV X-rays and  $n$ is the gas number density. The final term takes account of the effects of X-ray photoelectric absorption where $N_{22}$ is the the column density in units of $10^{22}$~cm$^{-2}$ \citep[In Cygnus A X-ray observations give an 
estimate of $N_{22}=20$ along the line of sight to the core,][]{young02}.  According to  the model of \citet[see their Fig.~3]{maloney96a}~ for  Cygnus~A at $r=80$~pc the gas fraction is  predominantly atomic ($>90$\%) for a density range $10^3<n<1.6\cdot 10^5$~cm$^{-3}$. The remaining $<10$\% of the gas is ionised at low densities ($n=10^3$~cm$^{-3}$) and becomes molecular at higher densities ($n\approx10^5$~cm$^{-3}$). Above densities of $3.3\cdot 10^5$~cm$^{-3}$ the gas is mostly molecular. 

The \citet{maloney96a} model can be used to predict the disk \HI and free-free absorption opacity as a function of gas density; comparison with observations can then constrain the density. In order to predict opacities we need to convert gas volume densities to column densities which 
requires estimates of the path length through the absorbing gas. Based on  our estimate of the disk thickness of $~20$~pc and the maximum disk inclination angle with respect to the line of sight (Sect.~4.1) we estimate the geometric path length through the disk  to be at  least 31~pc. This is a strict lower  limit and  we adopt $L_{\rm{geom}}=40$~pc as a more likely value. Note that if the gas exists in clouds the effective path length through absorbing gas  can be less than $L_{\rm{geom}}$ but it cannot be larger.  Using the above assumptions
we find that upper limits on free-free absorption toward the counter-jet do not give any useful gas density constraints. This is because, although the free-electron fraction increases at lower densities this is compensated for by a lower total gas column such that  the free-free opacity continuum absorption at 21cm 
wavelength against the counter-jet is less than 0.1 over all reasonable densities.

For modelling \HI absorption the relevant parameters given by \citet{maloney96a}  are the predicted  gas temperature and atomic fraction  versus  effective ionisation parameter $\xi_{\rm{eff}}$.   
At the radius of the absorbing \HI assuming  densities high enough to give a predominately atomic (and not ionised)  column, the increase of $T_{\rm{spin}}$ due to radiative excitation from the radio core is  negligible \citep[see e.g.][]{bahcall69}. Given this we assume the atomic gas is thermalised and that  its spin temperature ($T_{\rm{spin}}$)  equals the gas temperature  ($T$).  The predicted \HI line width in this 
cases is proportional to $N_{\rm{HI}} T^{-1} $  where $N_{\rm{HI}} $ is the \HI column density and $T$ the temperature. Both of the above quantities 
can be calculated as functions of the local density $n$.  The former quantity equals the product of the  total column density $N = n \cdot L_{\rm{geom}} $  (for a uniform filled
column) multiplied by the \HI abundance. This abundance is a function of $\xi_{\rm{eff}}$   \citep[calculated using Fig.~3 in][]{maloney96a} 
 which in turn is a function of $n$ via eq.~2 (where $ N_{22} = N/10^{22}$).  Likewise $T$ can be derived as a function of $n$, Êagain using 
Fig.~3 in \citep{maloney96a}.  Combining the $N_{\rm{HI}}$Ê  and  $ T $ dependencies on $n$ together we can predict $N_{\rm{HI}} T^{-1} $ 
and hence the expected \HI absorption linewidth
versus local density and then compare to observations. For densities $n>10^4$~cm$^{-3}$  the predicted \HI absorption is
 larger than observed, this can however easily be reconciled with observations 
if the \HI absorption is concentrated in clouds (see below).  At densities $n<10^4$~cm$^{-3}$  however there are no solutions that can provide
enough \HI  absorption to match what we observe, hence we can set a minimum density of  $10^4$~cm$^{-3}$  for our \HI absorbing gas.

For the gas density range $1.0\cdot 10^4$~cm$^{-3} < n<3.2\cdot 10^4$~cm$^{-3}$ there exist pure atomic phase gas solutions which fit the \HI observations. These solutions require the absorption to occur in clouds such that only a fraction $f_{\rm{LOS}}$ of a typical LOS passes through clouds. 
At the higher densities within this range the requirements on total gas column density are reduced  because gas/spin temperatures decline making hydrogen atoms more efficient at absorption. At the boundary in density  between having predominately atomic or molecular phase clouds (at density $4.6\cdot 10^4$~cm$^{-3}$) the spin temperature is $\sim 114$~K the cloud line of sight filling factor is $f_{\rm{LOS}}=0.032$ and $N=1.82\cdot 10^{23}$~cm$^{-2}$ which is comparable to the X-ray absorption column along the line of sight to the central engine. At higher densities the clouds become predominately molecular but the observed  \HI absorption  can still be fitted by similar total gas column densities. The reason is that although the atomic fraction declines rapidly with increasing density the temperature  also declines, this increases the efficiency of absorption per  hydrogen atom which almost exactly compensates for the reduced abundance. 

In summary we cannot on the basis of our \HI observations alone distinguish between models for the \HI absorbing gas phase which are primarily atomic or molecular. We can however set at minimum density ($n>10^4$~cm$^{-3}$) and a range for the total gas column density through the  \HI absorbing disk ($10^{23}$~cm$^{-2} < N< 10^{24}$~cm$^{-2}$). Estimating total gas masses is complicated for those solutions invoking clouds in that these solutions only fix the line of sight filling factor, converting to volume filling factors requires knowing the cloud size, on which we have no constraint. For the  minimum density solution where the gas column is continuous we can make a rough estimate of disk gas mass of  $M_{\rm{gas disk}}=1.0\cdot10^8$~\Msun ~within the inner 80~pc radius of the disk. For higher density solutions, both total required column densities and cloud filling factors decrease so that total gas mass requirements are significantly less. In all cases the total disk mass is much less than that of the central black hole \citep[$M_{\rm{BH}}=2.5\cdot10^9$~\Msun ,][]{tadhunter03} which therefore dominates the kinematics in the inner part of the nucleus that we observe.

 \subsection{Constraints on circumnuclear torus properties and relation to the \HI disk}

It is interesting that the estimated total gas column density through the \HI absorbing gas (see Section 4.2) is comparable to that estimated by X-ray photoelectric absorption toward the core of $2\cdot 10^{23}$~cm$^{-2}$ \citep[see][]{young02}.  Is it possible that most of this column density and most of the material that hides the quasar nucleus occurs as far out as $r=80$~pc?  An obvious problem with such a scenario is the absence of broad \HI absorption along the direct line of sight to the radio core  (peak opacity is $\tau_0 = 0.016 \pm 0,006$,  compared to $>1$  on the counter-jet,  see Sect.~3.3 and Fig.~\ref{opacity.pv}). A change in gas physical state with scale-height in the disk  while keeping almost constant column density seems unlikely.  If the high scale-height gas covering the core were still predominately atomic or molecular we find no density solutions that fit the low limit on \HI absorption seen. On the other hand if the high scale height gas were wholly ionised at this column density  it would strongly free-free absorb the radio core. Although there are  signs of free-free absorption at $r\le20$~pc \citep{krichbaum98} along the counter-jet  this absorption does not cover the core position itself.

It seems likely instead that most of the X-rays absorbing column in Cygnus~A and other hidden quasars occurs on scales $\ll 100$~pc in a compact circumnuclear {\sl torus}. The inner radius of such a torus is set by the dust sublimation radius, $r_{\rm{d}} = 0.4 \cdot (L_{\rm{bol}}\cdot10^{-45}$~erg$^{-1}$~s)$^{0.5}$~pc \citep{nenkova08} where $L_{\rm{bol}}$ is the AGN bolometric luminosity. For Cygnus~A $L_{\rm{bol}}$ is estimated to be $1.5 \cdot 10^{45}$~erg~s$^{-1}$ \citep{whysong04} giving $r_{\rm{d}}=0.5$~pc. It should be noted that the \citet{whysong04} mid-IR estimate of the bolometric luminosity is significantly lower (by factors $5-20$) than estimates based on the unabsorbed hard X-ray luminosity \citep[see][]{tadhunter03}, possibly because of uncorrected torus extinction at mid-IR wavelengths (C. Tadhunter, private communication). However, an increase by a factor of 20 in bolometric luminosity results in the sublimation radius increasing only to $r_{\rm{d}}=2.2$~pc, still significantly smaller than the radius of the \HI structure. The outer radius ($r_{\rm{out}}$) is probably not as cleanly defined as in some early depictions of doughnut-like tori \citep{padovani92} and may even be continuous with larger circumnuclear disks (see references in Sect.~1). Consistent with these more general geometries the term 'torus' can be thought of, if one prefers,  as an acronym for 'Thick/ toroidal Obscuration Required  by Unified Schemes' \citep{conway99,elitzur08} rather than referring to a specific fixed geometry.

Recently \citet{privon09} has fitted  the observed Spectral Energy  
Distribution (SED) of the Cygnus~A nucleus with a combined synchrotron  
jet, starburst and torus model. The torus fitting assumed the clumpy  
model of \citet{nenkova08b,nenkova08}. A wide range of solutions were  
obtained depending on assumptions about the disk inclination and  
synchrotron jet spectrum. Most plausible solutions however had cloud  
number density per unit volume declining as $r^{-q}$ with $q=1$ and ratios of outer to  
inner radius, $Y=30$ (giving $r_{\rm{out}}=18$~pc for Cygnus~A).  
Satisfyingly, the predicted total column density matches the X-ray  
absorption estimate within the errors. Despite the relatively large  
formal outer radius the concentration of clumps toward the inner edge   
means that most of the column density is concentrated at small radii.  
For a radial exponent of $q=1$ half of the total column density occurs  
within $\sqrt{Y} \cdot r_{\rm{in}}$ corresponding to 2.8~pc, i.e. on a  
much smaller scale than our observed \HI absorption.

The above results beg the question of relationship between the \HI absorbing disk and the circumnuclear torus in Cygnus A. The mass in the torus is very low \citep[$6\cdot 10^5$~\Msun , following][]{elitzur08} and the lifetime of large scale-height clouds within it is short \citep[due to intercloud collisions, see e.g.][]{krolik89}. Hence, both to fuel the quasar-like nucleus at the expected rate of $\sim 1$~\Msun yr$^{-1}$  and to replenish the torus clouds a much larger reservoir of material is required, which could reside in the \HI absorbing disk. The mass in such a disk (see Section 4.2) is sufficient to power the source for $10^{7}$ to $10^{8}$~yr and may form part of a 'feeding structure' which funnels gas from kilo-parsec scales into the central black hole. A feeding connection between disk and torus seems likely but the mechanisms of gas transport between the two scales is  unclear.

Pertinent to the feeding question is whether the torus and the \HI  
absorbing disk are a continuous structure with one gradually melding  
into the other, or if they are dynamically distinct and contain gas  
with very different physical conditions. The former possibility is  
motivated by the fitted torus outer radius $r_{\rm{out}}=18$~pc which  
is of the same order of magnitude as the radius of the \HI absorption  
($r=80$~pc). Furthermore the torus outer radius determined from SED 
fitting is   not well constrained by the data  and may extend beyond  
this limit. However, in such a continuous model for clouds with approximately fixed  
density the clouds would likely be in an increasingly molecular state over a  
projected radius from 1 to 70~pc as the excitation parameter decreased.
It is hard to see how the rapid  
gradient in \HI absorption seen along the counter-jet could be  
reproduced in such a case. Furthermore the observation of free-free  
absorption at intermediate positions \citep[$r<20$~pc,][]{krichbaum98}  
is inconsistent with such a continuous model. It seems more likely  
that the torus --- although ultimately fed from gas in the disk --- is  
a distinct structure, perhaps generated in an accretion disk wind as  
in the model of \citet{elitzur06}. Clearly, more work is required to  
understand how the observed components in Cygnus A and other radio  
galaxies of 100~pc scale  \HI absorbing disks, inner ionised gas and  
circumnuclear tori  are connected in self-consistent structures which  
both obscure and feed the central engine.

An additional remark concerning the disk/torus inter-relationship is that although in Cygnus~A the \HI absorbing disk likely does not contribute to 
the total absorbing column toward the central engine (as estimated from the X-ray observations) it could have  done so if we had observed Cygnus~A from a different direction.

Further progress on understanding the circumnuclear gas environment in  
Cygnus~A probably requires the reliable detection of molecular  
absorption and its imaging with VLBI. Single dish searches at  
commonly observed molecular transitions such as the ground and higher  
rotational transitions of CO \citep{barvainis94,salome03} have so far  
yielded only upper limits. In contrast a tentative detection of CO$+$  
in absorption using the IRAM 30m was reported by \citet{fuente00} with  
a centroid velocity and FWHM velocity width (170~\kms ) very similar  
to that of our broad \HI absorption. This detection has yet to be  
confirmed interferometrically (A. Fuente, private communication).  
Given the results in this paper, despite the
similarity in spectral shape, it seems unlikely that these  
observations are probing the same gas column as seen in  \HI  
absorption. The counter-jet which provides the background continuum  
against which the atomic hydrogen absorption is seen will have  
negligible flux density at millimetre wavelengths. Instead the line of  
sight probed at 3mm wavelength will probably lie at projected  
distances $<1$~pc from the central engine. While this is comparable to  
the scales on which the circumnuclear torus has its largest column  
density ($\sim 3$~pc) we would expect that the velocity dispersion of  
clouds in such a  geometrically thick torus would be comparable to the  
orbital velocity at this radius (i.e. 1900~\kms ), which is much  
broader than is observed. Similar considerations apply to the  
tentative VLBA observations of exited OH absorption reported by  
\citet{impellizzeri06} at projected radii  $<3$~pc  which only have  
observed width $\sim$100~\kms.

\subsection{The narrow absorption component}

While we argue that the broad absorption component is caused by gas rotating around the central black hole (Sect.~4.1)  the narrow ($<30$~\kms ~see Sect.~3)  absorption gas is likely to have a different origin. This narrow velocity component is significantly redshifted ($\sim 186$~\kms) ~with respect to the optical systemic velocity implying foreground gas moving inward toward the nucleus. This component is detected over the whole continuum region (i.e. jet, core and counterjet) where the observations are sensitive enough to detect an opacity of $\sim 0.1$ (see Sect.~3.2 and Fig.~\ref{opacity.pv}). A very similar opacity and centroid velocity is seen at the core position and one effective beam away on the jet side. The apparent increase in opacity at the velocity of the narrow absorption that is seen on the counter-jet side in  Fig.~\ref{opacity.pv} can be explained entirely by contamination by the broad velocity wings of the broad  velocity absorption. While we do not have many independent samples of the narrow velocity component opacity versus position  those we do have are consistent with a constant narrow component opacity of 0.1 over the whole VLBI radio source. Although the foreground distance from the narrow \HI absorption to the VLBI continuum source is not well constrained we suspect that it must be $>100$~pc given its narrow velocity width and the position stability of its velocity centroid; if the gas was closer one would expect tidal forces to widen the velocity width and increase velocity centroid variations.

The structure and origin of the narrow \HI absorption is unclear. Physically it could be related to a minor merging event, detected 400~pc South-West of the core \citep{canalizo03}. As a result of this interaction the narrow \HI component could arise in a tidal tail of gas that is moving towards the nucleus. According to galaxy merger simulations \citep{bournaud05} part of the progenitor's gas that gets expelled in tidal tails during a merging event will eventually fall back on the disk. The narrow component could also be connected to the giant infalling molecular cloud located 1.35~kpc to the North-West of the nucleus \citep{bellamy04}. However, the projected distances from the core are rather different such that a direct connection is not obvious. Both of the above gas components are relatively large ($>100$~pc) and thus would be consistent with having fairly constant \HI opacity over the whole VLBI structure.

Further progress with constraining the size and origin of the narrow \HI absorption system  might be made by making future sensitive (and high spectral resolution) e-MERLIN or EVLA observations to try to trace the narrow \HI absorption further along the jet.

\section{Summary}

We have presented VLBA \HI absorption data of the core region of Cygnus~A. \HI absorption is detected over  a linear scale of 95~pc, but is seen mainly along the counter-jet and on the nucleus. The integrated spectrum can be well-fitted by two Gaussian profiles suggesting a broad (FWHM$= 231\pm 21$~\kms ) and a narrow ($< 30$~\kms ) component. Modelling the data shows that the broad absorption occurs only against the counter-jet and not against the jet. Against the unresolved core the opacity is very low. The broad velocity component can be  explained by a circumnuclear disk which has its highest opacity 45~mas away  (in projection)  from the black hole. The narrow velocity component could be explained as an infalling tidal tail, presumably left from a past minor merging event.

The radius at which the broad  \HI absorption occurs can be constrained  with the help of limits on the orientation of the disk. We find a relatively narrow range of possible parameters, resulting in an estimated  radius of $\sim 80$~pc for the peak opacity, a disk scale height of about 20~pc and hence an opening angle of 14\degr . The offset between the centroid of the broad velocity component and the mean systemic velocity can be explained by a tilt of 21\degr ~of the disk axis compared to the jet axis.

Based on the derived geometry of the circumnuclear disk we derived physical properties of the \HI absorbing gas. We find the minimum gas density to be $n > 10^4$~cm$^{-3}$ with a spin temperature $T_{\rm{spin}}<740$~K. With the \HI observations alone we cannot distinguish between models for the \HI absorbing gas phase which are primarily atomic or molecular.  We can however set  a range for the total gas column density through the \HI absorbing disk of $10^{23}$~cm$^{-2}$~$< N < 10^{24}$~cm$^{-2}$. An upper limit on the gas mass within a radius of 80~pc is $M_{\rm{gas-disk}}=10^8$~\Msun , which is a factor 25 less than the black hole mass estimated by \citet{tadhunter03}.

The circumnuclear torus in Cygnus~A has an estimated fiducial radius $\sim3$~pc, which is a much smaller scale than our observed \HI absorption. The estimated mass in the torus clouds  ($6\cdot10^5$~\Msun ) is too low to power (alone) the source for $10^7-10^8$~yr. In contrast  the \HI disk has enough mass to feed the AGN and replenish the torus clouds.

Higher sensitivity, broader band observations are needed to study the properties of the atomic circumnuclear disk in more detail, to search for a connection with the torus and to investigate whether the torus contains a very 
broad (2000 \kms~ wide) \HI absorption component. Higher spectral resolution e-MERLIN or EVLA observations of the narrow absorption component are needed to try to trace absorption further out along the 
jet to constrain its physical size.


\begin{acknowledgements}

This research was supported by the EU Framework 6 Marie Curie Early Stage Training programme under contract number MEST-CT-2005-19669 ESTRELA. JC acknowledges financial support from the Swedish Science Research Council. We thank the referee Clive Tadhunter for his comments that helped to improve the paper. We wish to thank Joan Wrobel of NRAO for extensive help in setting up the phased VLA for VLBI observations at a non-standard observing frequency.

\end{acknowledgements}


\begin{thebibliography}{43}
\expandafter\ifx\csname natexlab\endcsname\relax\def\natexlab#1{#1}\fi

\bibitem[{{Antonucci}(1993)}]{antonucci93}
{Antonucci}, R. 1993, \araa, 31, 473

\bibitem[{{Bahcall} \& {Ekers}(1969)}]{bahcall69}
{Bahcall}, J.~N. \& {Ekers}, R.~D. 1969, \apj, 157, 1055

\bibitem[{{Barvainis} \& {Antonucci}(1994)}]{barvainis94}
{Barvainis}, R. \& {Antonucci}, R. 1994, \aj, 107, 1291

\bibitem[{{Bellamy} \& {Tadhunter}(2004)}]{bellamy04}
{Bellamy}, M.~J. \& {Tadhunter}, C.~N. 2004, \mnras, 353, 105

\bibitem[{{Bournaud} {et~al.}(2005){Bournaud}, {Jog}, \& {Combes}}]{bournaud05}
{Bournaud}, F., {Jog}, C.~J., \& {Combes}, F. 2005, \aap, 437, 69

\bibitem[{{Canalizo} {et~al.}(2003){Canalizo}, {Max}, {Whysong}, {Antonucci},
  \& {Dahm}}]{canalizo03}
{Canalizo}, G., {Max}, C., {Whysong}, D., {Antonucci}, R., \& {Dahm}, S.~E.
  2003, \apj, 597, 823

\bibitem[{{Carilli} {et~al.}(1991){Carilli}, {Perley}, {Dreher}, \&
  {Leahy}}]{carilli91}
{Carilli}, C.~L., {Perley}, R.~A., {Dreher}, J.~W., \& {Leahy}, J.~P. 1991,
  \apj, 383, 554

\bibitem[{{Conway}(1999)}]{conway99}
{Conway}, J.~E. 1999, in Astronomical Society of the Pacific Conference Series,
  Vol. 156, Highly Redshifted Radio Lines, ed. C.~L. {Carilli}, S.~J.~E.
  {Radford}, K.~M. {Menten}, \& G.~I. {Langston}, 259--+

\bibitem[{{Conway} \& {Blanco}(1995)}]{conway95}
{Conway}, J.~E. \& {Blanco}, P.~R. 1995, \apjl, 449, L131+

\bibitem[{{Elitzur}(2008)}]{elitzur08}
{Elitzur}, M. 2008, New Astronomy Review, 52, 274

\bibitem[{{Elitzur} \& {Shlosman}(2006)}]{elitzur06}
{Elitzur}, M. \& {Shlosman}, I. 2006, \apjl, 648, L101

\bibitem[{{Fanaroff} \& {Riley}(1974)}]{fanaroff74}
{Fanaroff}, B.~L. \& {Riley}, J.~M. 1974, \mnras, 167, 31P

\bibitem[{{Fuente} {et~al.}(2000){Fuente}, {Black}, {Mart{\'{\i}}n-Pintado},
  {Rodr{\'{\i}}guez-Franco}, {Garc{\'{\i}}a-Burillo}, {Planesas}, \&
  {Lindholm}}]{fuente00}
{Fuente}, A., {Black}, J.~H., {Mart{\'{\i}}n-Pintado}, J., {et~al.} 2000,
  \apjl, 545, L113

\bibitem[{{Hardcastle} {et~al.}(2009){Hardcastle}, {Evans}, \&
  {Croston}}]{hardcastle09}
{Hardcastle}, M.~J., {Evans}, D.~A., \& {Croston}, J.~H. 2009, \mnras, 396,
  1929

\bibitem[{{Hicks} {et~al.}(2009){Hicks}, {Davies}, {Malkan}, {Genzel},
  {Tacconi}, {S{\'a}nchez}, \& {Sternberg}}]{hicks09}
{Hicks}, E.~K.~S., {Davies}, R.~I., {Malkan}, M.~A., {et~al.} 2009, \apj, 696,
  448

\bibitem[{{Impellizzeri} {et~al.}(2006){Impellizzeri}, {Roy}, \&
  {Henkel}}]{impellizzeri06}
{Impellizzeri}, V., {Roy}, A.~L., \& {Henkel}, C. 2006, in Proceedings of the
  8th European VLBI Network Symposium

\bibitem[{{Jackson} {et~al.}(1998){Jackson}, {Tadhunter}, \&
  {Sparks}}]{jackson98}
{Jackson}, N., {Tadhunter}, C., \& {Sparks}, W.~B. 1998, \mnras, 301, 131

\bibitem[{{Jaffe} {et~al.}(2004){Jaffe}, {Meisenheimer}, {R{\"o}ttgering},
  {Leinert}, {Richichi}, {Chesneau}, {Fraix-Burnet}, {Glazenborg-Kluttig},
  {Granato}, {Graser}, {Heijligers}, {K{\"o}hler}, {Malbet}, {Miley},
  {Paresce}, {Pel}, {Perrin}, {Przygodda}, {Schoeller}, {Sol}, {Waters},
  {Weigelt}, {Woillez}, \& {de Zeeuw}}]{jaffe04}
{Jaffe}, W., {Meisenheimer}, K., {R{\"o}ttgering}, H.~J.~A., {et~al.} 2004,
  \nat, 429, 47

\bibitem[{{Jones} {et~al.}(1996){Jones}, {Tingay}, {Murphy}, {Meier},
  {Jauncey}, {Reynolds}, {Tzioumis}, {Preston}, {McCulloch}, {Costa},
  {Kemball}, {Nicolson}, {Quick}, {King}, {Lovell}, {Clay}, {Ferris}, {Gough},
  {Sinclair}, {Ellingsen}, {Edwards}, {Jones}, {van Ommen}, {Harbison}, \&
  {Migenes}}]{jones96}
{Jones}, D.~L., {Tingay}, S.~J., {Murphy}, D.~W., {et~al.} 1996, \apjl, 466,
  L63+

\bibitem[{{Jones} {et~al.}(2001){Jones}, {Wehrle}, {Piner}, \&
  {Meier}}]{jones01}
{Jones}, D.~L., {Wehrle}, A.~E., {Piner}, B.~G., \& {Meier}, D.~L. 2001, \apj,
  553, 968

\bibitem[{{Krichbaum} {et~al.}(1998){Krichbaum}, {Alef}, {Witzel}, {Zensus},
  {Booth}, {Greve}, \& {Rogers}}]{krichbaum98}
{Krichbaum}, T.~P., {Alef}, W., {Witzel}, A., {et~al.} 1998, \aap, 329, 873

\bibitem[{{Krolik} \& {Begelman}(1988)}]{krolik88}
{Krolik}, J.~H. \& {Begelman}, M.~C. 1988, \apj, 329, 702

\bibitem[{{Krolik} \& {Lepp}(1989)}]{krolik89}
{Krolik}, J.~H. \& {Lepp}, S. 1989, \apj, 347, 179

\bibitem[{{Lo}(2005)}]{lo05}
{Lo}, K.~Y. 2005, \araa, 43, 625

\bibitem[{{Maloney} {et~al.}(1996){Maloney}, {Hollenbach}, \&
  {Tielens}}]{maloney96a}
{Maloney}, P.~R., {Hollenbach}, D.~J., \& {Tielens}, A.~G.~G.~M. 1996, \apj,
  466, 561

\bibitem[{{Morganti} {et~al.}(2008){Morganti}, {Oosterloo}, {Struve}, \&
  {Saripalli}}]{morganti08}
{Morganti}, R., {Oosterloo}, T., {Struve}, C., \& {Saripalli}, L. 2008, \aap,
  485, L5

\bibitem[{{Nenkova} {et~al.}(2008{\natexlab{a}}){Nenkova}, {Sirocky},
  {Ivezi{\'c}}, \& {Elitzur}}]{nenkova08b}
{Nenkova}, M., {Sirocky}, M.~M., {Ivezi{\'c}}, {\v Z}., \& {Elitzur}, M.
  2008{\natexlab{a}}, \apj, 685, 147

\bibitem[{{Nenkova} {et~al.}(2008{\natexlab{b}}){Nenkova}, {Sirocky},
  {Nikutta}, {Ivezi{\'c}}, \& {Elitzur}}]{nenkova08}
{Nenkova}, M., {Sirocky}, M.~M., {Nikutta}, R., {Ivezi{\'c}}, {\v Z}., \&
  {Elitzur}, M. 2008{\natexlab{b}}, \apj, 685, 160

\bibitem[{{Ogle} {et~al.}(1997){Ogle}, {Cohen}, {Miller}, {Tran}, {Fosbury}, \&
  {Goodrich}}]{ogle97}
{Ogle}, P.~M., {Cohen}, M.~H., {Miller}, J.~S., {et~al.} 1997, \apjl, 482, L37+

\bibitem[{{Padovani} \& {Urry}(1992)}]{padovani92}
{Padovani}, P. \& {Urry}, C.~M. 1992, \apj, 387, 449

\bibitem[{{Peck} \& {Taylor}(2001)}]{peck01}
{Peck}, A.~B. \& {Taylor}, G.~B. 2001, \apjl, 554, L147

\bibitem[{{Privon}(2009)}]{privon09}
{Privon}, G.~C. 2009, ArXiv e-prints

\bibitem[{{Salom{\'e}} \& {Combes}(2003)}]{salome03}
{Salom{\'e}}, P. \& {Combes}, F. 2003, \aap, 412, 657

\bibitem[{{Schinnerer} {et~al.}(2000){Schinnerer}, {Eckart}, {Tacconi},
  {Genzel}, \& {Downes}}]{schinnerer00}
{Schinnerer}, E., {Eckart}, A., {Tacconi}, L.~J., {Genzel}, R., \& {Downes}, D.
  2000, \apj, 533, 850

\bibitem[{{Tadhunter}(2008)}]{tadhunter08}
{Tadhunter}, C. 2008, New Astronomy Review, 52, 227

\bibitem[{{Tadhunter} {et~al.}(2003){Tadhunter}, {Marconi}, {Axon}, {Wills},
  {Robinson}, \& {Jackson}}]{tadhunter03}
{Tadhunter}, C., {Marconi}, A., {Axon}, D., {et~al.} 2003, \mnras, 342, 861

\bibitem[{{Tadhunter} {et~al.}(1999){Tadhunter}, {Packham}, {Axon}, {Jackson},
  {Hough}, {Robinson}, {Young}, \& {Sparks}}]{tadhunter99}
{Tadhunter}, C.~N., {Packham}, C., {Axon}, D.~J., {et~al.} 1999, \apjl, 512,
  L91

\bibitem[{{Taylor}(1996)}]{taylor96}
{Taylor}, G.~B. 1996, \apj, 470, 394

\bibitem[{{Tristram} {et~al.}(2009){Tristram}, {Raban}, {Meisenheimer},
  {Jaffe}, {R{\"o}ttgering}, {Burtscher}, {Cotton}, {Graser}, {Henning},
  {Leinert}, {Lopez}, {Morel}, {Perrin}, \& {Wittkowski}}]{tristram09}
{Tristram}, K.~R.~W., {Raban}, D., {Meisenheimer}, K., {et~al.} 2009, \aap,
  502, 67

\bibitem[{{van Langevelde} {et~al.}(2000){van Langevelde}, {Pihlstr{\"o}m},
  {Conway}, {Jaffe}, \& {Schilizzi}}]{vanlangevelde00}
{van Langevelde}, H.~J., {Pihlstr{\"o}m}, Y.~M., {Conway}, J.~E., {Jaffe}, W.,
  \& {Schilizzi}, R.~T. 2000, \aap, 354, L45

\bibitem[{{Vermeulen} {et~al.}(1994){Vermeulen}, {Readhead}, \&
  {Backer}}]{vermeulen94}
{Vermeulen}, R.~C., {Readhead}, A.~C.~S., \& {Backer}, D.~C. 1994, \apjl, 430,
  L41

\bibitem[{{Whysong} \& {Antonucci}(2004)}]{whysong04}
{Whysong}, D. \& {Antonucci}, R. 2004, \apj, 602, 116

\bibitem[{{Young} {et~al.}(2002){Young}, {Wilson}, {Terashima}, {Arnaud}, \&
  {Smith}}]{young02}
{Young}, A.~J., {Wilson}, A.~S., {Terashima}, Y., {Arnaud}, K.~A., \& {Smith},
  D.~A. 2002, \apj, 564, 176

\end{thebibliography}
\end{document}